\pdfoutput=1

\documentclass[journal]{IEEEtran}

\ifCLASSINFOpdf
   \usepackage[pdftex]{graphicx}
   \DeclareGraphicsExtensions{.pdf,.jpeg,.png}
\else
  \usepackage[dvips]{graphicx}
  \DeclareGraphicsExtensions{.eps}
\fi
\usepackage{multirow}

\begin{document}
%
\title{Deep convolutional neural networks for predominant instrument recognition in polyphonic music}
%
%
%

\author{Yoonchang~Han, Jaehun Kim, and~Kyogu~Lee,~\IEEEmembership{Senior Member,~IEEE}
\thanks{Y. Han, J. Kim, and K. Lee are with the Music and Audio Research Group, Graduate School of Convergence Science and Technology, Seoul National University, Seoul 08826, Republic of Korea, e-mail: (yoonchanghan@snu.ac.kr, eldrin@snu.ac.kr, kglee@snu.ac.kr).}
\thanks{K. Lee is also with the Advanced Institutes of Convergence Technology, Suwon, Republic of Korea}
\thanks{Manuscript received \today; revised \today.}}

%
%

\markboth{Journal of \LaTeX\ Class Files,~Vol.~14, No.~8, May~2016}%
{Shell \MakeLowercase{\textit{et al.}}: Bare Demo of IEEEtran.cls for IEEE Journals}
%


\maketitle

\begin{abstract}
Identifying musical instruments in polyphonic music recordings is a challenging but important problem in the field of music information retrieval. It enables music search by instrument, helps recognize musical genres, or can make music transcription easier and more accurate. In this paper, we present a convolutional neural network framework for predominant instrument recognition in real-world polyphonic music. We train our network from fixed-length music excerpts with a single-labeled predominant instrument and estimate an arbitrary number of predominant instruments from an audio signal with a variable length. To obtain the audio-excerpt-wise result, we aggregate multiple outputs from sliding windows over the test audio. In doing so, we investigated two different aggregation methods: one takes the average for each instrument and the other takes the instrument-wise sum followed by normalization. In addition, we conducted extensive experiments on several important factors that affect the performance, including analysis window size, identification threshold, and activation functions for neural networks to find the optimal set of parameters. Using a dataset of 10k audio excerpts from 11 instruments for evaluation, we found that convolutional neural networks are more robust than conventional methods that exploit spectral features and source separation with support vector machines. Experimental results showed that the proposed convolutional network architecture obtained an F1 measure of 0.602 for micro and 0.503 for macro, respectively, achieving 19.6\% and 16.4\% in performance improvement compared with other state-of-the-art algorithms.
\end{abstract}


\ifCLASSOPTIONpeerreview
\begin{center} \bfseries EDICS Category: 3-BBND \end{center}
\fi
%
\IEEEpeerreviewmaketitle

\section{Introduction}
\IEEEPARstart{M}{usic} can be said to be built by the interplay of various instruments. A human can easily identify what instruments are used in a music, but it is still a difficult task for a computer to automatically recognize them. This is mainly because music in the real world is mostly polyphonic, which makes extraction of information from an audio highly challenging. Furthermore, instrument sounds in the real world vary in many ways such as for timbre, quality, and playing style, which makes identification of the musical instrument even harder.

In the music information retrieval (MIR) field, it is highly desirable to know what instruments are used in an audio sample. First of all, instrument information per se is an important and useful information for users, and it can be included in the audio tags. There is a huge demand for music search owing to the increasing number of music files in digital format. Unlike text search, it is difficult to search for music because input queries are usually in text format. If an instrument information is included in the tags, it allows people to search for music with the specific instrument they want. In addition, the obtained instrument information can be used for various audio/music applications. For instance, more instrument-specific and tailored audio equalization can be applied to the music; moreover, a music recommendation system can reflect the preference of users for musical instruments. Furthermore, it can also be used to enhance the performance of other MIR tasks. For example, knowing the number and type of the instrument would significantly improve the performance of source separation and automatic music transcription; it would also be helpful for identifying the genre of the music.

Instrument recognition can be performed in various forms. Hence, the term ``instrument recognition'' or ``instrument identification'' might indicate several different research topics. For instance, many of the related works focus on studio-recorded isolated notes. To name a few, Eronen used cepstral coefficients and temporal features to classify 30 orchestral instruments with several articulation styles and achieved a classification accuracy of 95\% for instrument family level and about 81\% for individual instruments \cite{eronen2000musical}. Diment et al. used a modified group delay feature that incorporates phase information together with mel-frequency cepstral coefficients (MFCCs) and achieved a classification accuracy of about 71\% for 22 instruments \cite{diment2013modified}. Yu et al. applied sparse coding on cepstrum with temporal sum-pooling and achieved an F-measure of about 96\% for classifying 50 instruments \cite{yu2014sparse}. They also reported their classification result on a multi-source database, which was about 66\%. 

Some previous works such as Krishna and Sreenivas \cite{krishna2004music} experimented with a classification for solo phrases rather than for isolated notes. They proposed line spectral frequencies (LSF) with a Gaussian mixture model (GMM) and achieved an accuracy of about 77\% for instrument family and 84\% for 14 individual instruments. Moreover, Essid et al. \cite{essid2004musical} reported that a classification system with MFCCs and GMM along with principal components analysis (PCA) achieved an overall recognition accuracy of about 67\% on solo phrases with five instruments.

More recent works deal with polyphonic sound, which is closer to real-world music than to monophonic sound. In the case of polyphonic sound, a number of research studies used synthesized polyphonic audio from studio-recorded single tones. Heittola et al. \cite{heittola2009musical} used a non-negative matrix factorization (NMF)-based source-filter model with MFCCs and GMM for synthesized polyphonic sound and achieved a recognition rate of 59\% for six polyphonic notes randomly generated from 19 instruments. Kitahara et al. \cite{kitahara2007instrument} used various spectral, temporal, and modulation features with PCA and linear discriminant analysis (LDA) for classification. They reported that, using feature weighting and musical context, recognition rates were about 84\% for a duo, 78\% for a trio, and 72\% for a quartet. Duan et al. \cite{duan2014novel} proposed the uniform discrete cepstrum (UDC) and mel-scale UDC (MUDC) as a spectral representation with a radial basis function (RBF) kernel support vector machine (SVM) to classify 13 types of Western instruments. The classification accuracy of randomly mixed chords of two and six polyphonic notes, generated using isolated note samples from the RWC musical instrument sound database \cite{goto2003rwc}, was around 37\% for two polyphony notes and 25\% for six polyphony notes.

As shown above, most of the previous works focused on the identification of the instrument sounds in clean solo tones or phrases. More recent research studies on polyphonic sounds are closer to the real-world situation, but artificially produced polyphonic music is still far from professionally produced music. Real-world music has many other factors that affect the recognition performance. For instance, it might have a highly different timbre, depending on the genre and style of the performance. In addition, an audio file might differ in quality to a great extent, depending on the recording and production environments.

In this paper, we investigate a method for predominant instrument recognition in professionally produced Western music recordings. We utilize convolutional neural networks (ConvNets) to learn the spectral characteristics of the music recordings with 11 musical instruments and perform instrument identification on polyphonic music excerpts. The major contributions of the work presented in this paper are as follows.

\par \vspace{\baselineskip}
\hangindent=0.7cm 
1. We present the ConvNet architecture for predominant musical instrument identification where the training data are single labeled and the target data are multi-labeled with an unknown number of classes existing in the data.
\par \vspace{\baselineskip}
\hangindent=0.7cm 
2. We introduce a new method to aggregate the outputs of ConvNets from short-time sliding windows to find the predominant instruments in a music excerpt with variable length, where the conventional method of majority vote often fails.
\par \vspace{\baselineskip}
\hangindent=0.7cm 
3. We conduct an extensive experiment on activation function for the neurons used in ConvNets, which can cause a huge impact on the identification result.
\par \vspace{\baselineskip}

The remainder of the paper is organized as follows. In section II, we introduce emerging deep neural network techniques in the MIR field. Next, the system architecture section includes audio preprocessing, the proposed network architecture with detailed training configuration, and an explanation of various activation functions used for the experiment. Section IV, the evaluation section, contains information about the dataset, testing configuration including aggregation strategy, and our evaluation scheme. Then, we illustrate the performance of the proposed ConvNet in section V, the Results section, with an analysis of the effects of activation function, analysis window size, aggregation strategy, and identification threshold, and with an instrument-wise analysis. Moreover, we present a qualitative analysis based on the visualization of the ConvNet's intermediate outputs to understand how the network captured the pattern from the input data. Finally, we conclude the paper in section VI.

\section{Proliferation of Deep Neural Networks in Music Information Retrieval}
The ability of traditional machine learning approaches was limited in terms of processing input data in their raw form. Hence, usually the input for the learning system, typically a classifier, has to be a hand-crafted feature representation, which requires extensive domain knowledge and a careful engineering process. However, it is getting more common to design the system to automatically discover the higher-level representation from the raw data by stacking several layers of nonlinear modules, which is called deep learning \cite{lecun2015deep}. Recently, deep learning techniques have been widely used across a number of domains owing to their superior performance. A basic architecture of deep learning is called deep neural network (DNN), which is a feedforward network with multiple hidden layers of artificial neurons. DNN-based approaches have outperformed previous state-of-the-art methods in speech applications such as phone recognition, large-vocabulary speech recognition, multi-lingual speech recognition, and noise-robust speech recognition \cite{deng2014deep}.

There are many variants and modified architectures of deep learning, depending on the target task. Especially, recurrent neural networks (RNNs) and ConvNets have recently shown remarkable results for various multimedia information retrieval tasks. RNNs are highly powerful approaches for sequential inputs as their recurrent architecture enables their hidden units to implicitly maintain the information about the past elements of the sequence. Since languages natively contain sequential information, it is widely applied to handle text characters or spoken language. It has been reported that RNNs have shown a successful result on language modeling \cite{mikolov2010recurrent} and spoken language understanding \cite{mesnil2013investigation, yao2013recurrent}.

\begin{figure*}[!t]
  \centering
  \includegraphics[keepaspectratio, width=6.2in]{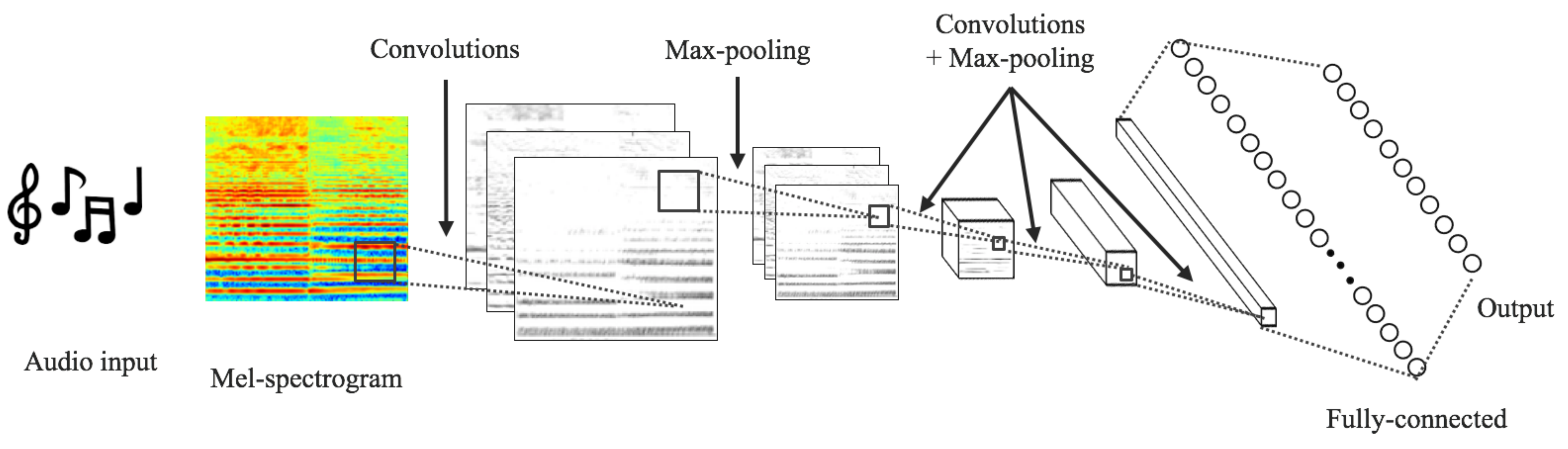}
  \caption{Schematic of the proposed ConvNet containing 4 times repeated double convolution layers followed by max-pooling. The last max-pooling layer performs global max-pooling, then it is fed to a fully connected layer followed by 11 sigmoid outputs.}
\label{fig:figure_sysArch}
\end{figure*}

On the other hand, ConvNet is useful for data with local groups of values that are highly correlated, forming distinctive local characteristics that might appear at different parts of the array \cite{lecun2015deep}. Hence, it is one of the most popular approaches recently in the image processing area such as handwritten digit recognition \cite{lecun1995learning, calderon2003handwritten,niu2012novel} for the MNIST dataset and image tagging \cite{krizhevsky2012imagenet,ngiam2010tiled} for the CIFAR-10 dataset. In addition, it has been reported that it has outperformed state-of-the-art approaches for several computer vision benchmark tasks such as object detection, semantic segmentation, and category-level object recognition \cite{deng2014deep}, and also for speech-recognition tasks \cite{hinton2012deep}.

The time-frequency representation of a music signal is composed of harmonics from various musical instruments and a human voice. Each musical instrument produces a unique timbre with different playing styles, and this type of spectral characteristics in music signal might appear in a different location in time and frequency as in the image. ConvNets are usually composed of many convolutional layers, and inserting a pooling layer between convolutional layers allows the network to work at different time scales and introduces translation invariance with robustness against local distortions. These hierarchical network structures of ConvNets are highly suitable for representing music audio, because music tends to present a hierarchical structure in time and different features of the music might be more salient at different time scales \cite{hamel2011temporal}. 

Hence, although ConvNets have been a more commonly used technique in image processing, there are an increasing number of attempts to apply ConvNets for music signal. It has been reported that ConvNet has outperformed previous state-of-the-art approaches for various MIR tasks such as onset detection \cite{schluter2014improved}, automatic chord recognition \cite{humphrey2012rethinking, boulanger2013audio}, and music structure/boundary analysis \cite{ullrich2014boundary, grill2015music}.

An attempt to apply ConvNets for musical instrument identification can be found in the recent report from Park et al. \cite{park2015musical} and Li et al. \cite{li2015automatic}, although it is still an ongoing work and is not a predominant instrument recognition method; hence, there are no other instruments but only target instrument sounds exist. Our research differs from \cite{park2015musical} because we deal with polyphonic music, while their work is based on the studio recording of single tones. In addition, our research also differs from \cite{li2015automatic} because we use single-label data for training and estimate multi-label data, while they used multi-label data from the training phase. Moreover, they focused on an end-to-end approach, which is promising in that using raw audio signals makes the system rely less on domain knowledge and preprocessing, but usually it shows a slightly lower performance than using spectral input such as mel-spectrogram in recent papers \cite{hoshen2015speech,palaz2013estimating}.

\section{System architecture}

\subsection{Audio Preprocessing}
The convolutional neural network is one of the representation learning methods that allow a machine to be fed with raw data and to automatically discover the representations needed for classification or detection \cite{lecun2015deep}. However, appropriate preprocessing of input data is still an important issue to improve the performance of the system.

In the first preprocessing step, the stereo input audio is converted to mono by taking the mean of the left and right channels, and then it is downsampled to 22,050 Hz from the original 44,100 Hz of sampling frequency. This allows us to use frequencies up to 11,025 Hz, the Nyquist frequency, and it is sufficient to cover most of the harmonics generated by musical instruments while removing noises possibly included in the frequencies above this range. Moreover, all audios are normalized by dividing the time-domain signal with its maximum value. Then, this downsampled time-domain waveform is converted to a time-frequency representation using short-time Fourier transform (STFT) with 1024 samples for the window size (approx. 46 ms) and 512 samples of the hop size (approx. 23 ms).

Next, the linear frequency scale-obtained spectrogram is converted to a mel-scale. We use 128 for the number of mel-frequency bins, following the representation learning papers on music annotation by Nam et al. \cite{nam2012learning} and Hamel et al. \cite{hamel2011temporal}, which is a reasonable setting that sufficiently preserves the harmonic characteristics of the music while greatly reducing the dimensionality of the input data. Finally, the magnitude of the obtained mel-frequency spectrogram is compressed with a natural logarithm.

\subsection{Network Architecture}
ConvNets can be seen as a combination of feature extractor and the classifier. Our ConvNet architecture generally follows a popular AlexNet \cite{krizhevsky2012imagenet} and VGGNet \cite{simonyan2014very} structure, which contains very deep architecture using repeated several convolution layers followed by max-pooling, as shown in Figure \ref{fig:figure_sysArch}. This method of using smaller receptive window size and smaller stride for ConvNet is becoming highly common especially in the computer vision field such as in the study from Zeiler and Fergus \cite{zeiler2014visualizing} and Sermanet et al. \cite{sermanet2013overfeat}, which has shown superior performance in ILSVRC-2013. 

Although the general architecture style is similar to that of other successful ConvNets in the image processing area, the proposed ConvNet is designed according to our input data. We use filters with a very small $3\times3$ receptive field, with a fixed stride size of 1, and spatial abstraction is done by max-pooling with a size of $3\times3$ and a stride size of 1.

In Table \ref{tab:convnet_arch}, we illustrate the detailed ConvNet architecture with the input size in each layer with parameter values except the zero-padding process. The input for each convolution layer is zero-padded with $1\times1$ to preserve the spatial resolution regardless of input window size, and we increase the number of channels for the convolution layer by a factor of 2 after every two convolution layers, starting from 32 up to 256.

In the last max-pooling layer after the eight convolutional layers, we perform global max-pooling followed by one fully connected layer. Recently, it has been reported that the use of global average pooling without a fully connected layer before a classifier layer is less prone to overfitting and shows better performance for image processing datasets such as CIFAR-10 and MNIST \cite{lin2013network}. However, our empirical experiment found that global average pooling slightly decreases the performance and that global max-pooling followed by a fully connected layer works better for our task.

Finally, the last classifier layer is the sigmoid layer. It is common to use a softmax layer when there is only one target label, but our system must be able to handle multiple instruments present at the same time, and, thus, a sigmoid output is used.

\begin{table}[!t]
\renewcommand{\arraystretch}{1.3}
\caption{Proposed ConvNet Structure. The Input Size Demonstrated in This Table Is for an Analysis Window Size of 1 Second (Number of filters $\times$ time $\times$ frequency). The Activation Function Is Followed by Each Convolutional Layer and a Fully Connected Layer. The Input of Each Convolution Layer Is Zero-Padded with 1 $\times$ 1, But Is Not Shown for Brevity.}
\centering
\begin{tabular}{l l}
    \hline
    \hline
    Input size & Description\\
    \hline
    1 $\times$ 43 $\times$ 128 & mel-spectrogram\\
    32 $\times$ 45 $\times$ 130 & 3 $\times$ 3 convolution, 32 filters\\
    32 $\times$ 47 $\times$ 132 & 3 $\times$ 3 convolution, 32 filters\\
    32 $\times$ 15 $\times$ 44 & 3 $\times$ 3 max-pooling\\
    32 $\times$ 15 $\times$ 44 & dropout (0.25)\\
    64 $\times$ 17 $\times$ 46 & 3 $\times$ 3 convolution, 64 filters\\
    64 $\times$ 19 $\times$ 48 & 3 $\times$ 3 convolution, 64 filters\\
    64 $\times$ 6 $\times$ 16 & 3 $\times$ 3 max-pooling\\
    64 $\times$ 6 $\times$ 16 & dropout (0.25)\\
    128 $\times$ 8 $\times$ 18 & 3 $\times$ 3 convolution, 128 filters\\
    128 $\times$ 10 $\times$ 20 & 3 $\times$ 3 convolution, 128 filters\\
    128 $\times$ 3 $\times$ 6 & 3 $\times$ 3 max-pooling\\
    128 $\times$ 3 $\times$ 6 & dropout (0.25)\\
    256 $\times$ 5 $\times$ 8 & 3 $\times$ 3 convolution, 256 filters\\
    256 $\times$ 7 $\times$ 10 & 3 $\times$ 3 convolution, 256 filters\\
    256 $\times$ 1 $\times$ 1 & global max-pooling\\
    1024 & flattened and fully connected\\
    1024 & dropout (0.50)\\
    11 & sigmoid\\
    \hline
    \hline
\end {tabular}
\label{tab:convnet_arch}
\end{table}

\subsection{Training Configuration}

The training was done by optimizing the categorical cross-entropy between predictions and targets. We used Adam \cite{kingma2014adam} as an optimizer with a learning rate of 0.001, and the mini-batch size was set to 128. To accelerate the learning process with parallelization, we used a GTX 970 GPU, which has 4GB of memory.

The training was regularized using dropout with a rate of 0.25 after each max-pooling layer. Dropout is a technique that prevents the overfitting of units to the training data by randomly dropping some units from the neural network during the training phase \cite{srivastava2014dropout}. Furthermore, we added dropout after a fully connected layer as well with a rate of 0.5 since a fully connected layer easily suffers from overfitting.

In addition, we conducted an experiment with various time resolutions to find the optimal analysis size. As our training data were a fixed 3-s audio, we performed the training with 3.0, 1.5, 1.0, and 0.5 s by dividing the training audio and used the same label for each divided chunk. The audio was divided without overlap for training as it affects the validation loss used for the early stopping. Fifteen percent of the training data were randomly selected and used as a validation set, and the training was stopped when the validation loss did not decrease for more than two epochs.

The initialization of the network weights is another important issue as it can lead to an unstable learning process, especially for a very deep network. We used a uniform distribution with zero biases for both convolutional and fully connected layers following Glorot and Bengio \cite{glorot2010understanding}.

\subsection{Activation Function}

The activation function is followed by each convolutional layer and fully connected layer. In this section, we introduce several activation functions used in the experiment for the comparison.

The traditional way to model the activation of a neuron is by using a hyperbolic tangent (tanh) or sigmoid function. However, non-saturating nonlinearities such as the rectified linear unit (ReLU) allow much faster learning than these saturating nonlinearities, particularly for models that are trained on large datasets \cite{krizhevsky2012imagenet}. Moreover, a number of works have shown that the performance of ReLU is better than that of sigmoid and tanh activation \cite{gu2015recent}. Thus, most of the modern studies on ConvNets use ReLU to model the output of the neurons \cite{li2015automatic,simonyan2014very,zeiler2014visualizing,sermanet2013overfeat}.

ReLU was first introduced by Nair and Hinton in their work on restricted Boltzmann machines \cite{nair2010rectified}. The ReLU activation function is defined as

\begin{equation}
y_i = max(0,z_i)
\end{equation}

\noindent where  $z_i$ is the input of the $i$th channel. ReLU simply suppresses the whole negative part to zero while retaining the positive part. 
Recently, there have been several modified versions of ReLU introduced to improve the performance further. First, leaky-ReLU (LReLU), introduced by Mass et al. \cite{maas2013rectifier}, compresses the negative part rather than make it all zero, which might cause some initially inactive units to remain inactive. It is defined as

\begin{equation}
y_i = \left\{
  \begin{array}{lr}
    z_i  &  z_i \ge 0\\
    \alpha z_i  &  z_i < 0
  \end{array}
\right.
\end{equation}

\noindent where  $\alpha$ is a parameter between 0 and 1 to give a small gradient in the negative part. Second, parametric ReLU (PReLU), introduced by He et al. \cite{he2015delving}, is basically similar to LReLU in that it compresses the negative part. However, PReLU automatically learns the parameter for the negative gradient, unlike LReLU. It is defined as
\begin{equation}
y_i = \left\{
  \begin{array}{lr}
    z_i  &  z_i \ge 0\\
    \alpha_iz_i  &  z_i < 0
  \end{array}
\right.
\end{equation}

\noindent where $\alpha_i$ is the learned parameters for the  $i$th channel.

The choice of activation function considerably influences the identification performance. It is difficult to say which specific activation function always performs the best because it highly depends on the parameter setting and the input data. For instance, an empirical evaluation of the ConvNet activation functions from Xu et al. \cite{xu2015empirical} reported that the performance of LReLU is better than those of ReLU and PReLU, but sometimes it is worse than that of basic ReLU, depending on the dataset and the value for $\alpha$. Moreover, most of the works regarding activation function are on the image classification task, not on the audio processing domain.

Hence, we empirically evaluated several activation functions explained above such as tanh, ReLU, LReLU, and PReLU to find the most suitable activation function for our task. For LReLU, very leaky ReLU ($\alpha$ = 0.33) and normal leaky ReLU ($\alpha$ = 0.01) were used, because it has been reported that the performance of LReLU considerably differs depending on  the value and that very leaky ReLU works better \cite{xu2015empirical}.

We used separate test audio data from the IRMAS dataset, which were not used for the training. First, a sliding window was used to analyze the input test audio, which was of the same size as the analysis window in the training phase. The hop size of the sliding window was set to half of the window size. Then, we aggregated the sigmoid outputs from the sliding windows by summing all outputs class-wise to obtain a total amount of activation for each instrument. These 11 summed sigmoid activations were then normalized to be in a range between 0 and 1 by dividing all with the maximum activation. 

\section{Evaluation}

\subsection{IRMAS Dataset}
The IRMAS dataset includes musical audio excerpts with annotations of the predominant instruments present and is intended to be used for the automatic identification of the predominant instruments in the music. This dataset was used in the paper on predominant instrument classification by Bosch et al. \cite{bosch2012comparison} and includes music from various decades from the past century, hence differing in audio quality to a great extent. In addition, the dataset covers a wide variability in musical instrument types, articulations, recording and production styles, and performers.

The dataset is divided into training and testing data, and all audio files are in 16-bit stereo wave with 44,100 Hz of sampling rate. The training data consisted of 6705 audio files with excerpts of 3 s from more than 2000 distinct recordings. Two subjects were paid to obtain the data for 11 pitched instruments, as shown in Table \ref{tab:instruments} from selected music tracks, with the objective of extracting music excerpts that contain a continuous presence of a single predominant instrument.

On the other hand, the testing data consisted of 2874 audio files with lengths between 5 s and 20 s, and no tracks from the training data were included. Unlike the training data, the testing data contained one or more predominant target instruments. Hence, the total number of training labels was identical to the number of audio files, but the number of testing labels was more than the number of testing audio files as the latter are multi-label. For both the training and the testing dataset, other musical instruments such as percussion and bass were not included in the annotation even if they exist in the music excerpts.

\begin{table}[!t]
\renewcommand{\arraystretch}{1.3}
\caption{List of Musical Instruments Used in the Experiment with Their Abbreviations, and the Number of Labels of the Training and Testing Audio.}
\centering
\begin{tabular}{l l l l}
    \hline
    \hline
    Instruments & Abbreviations & Training (n) & Testing (n)\\
    \hline
    Cello & cel & 388 & 111\\
    Clarinet & cla & 505 & 62\\
    Flute & flu & 451 & 163\\
    Acoustic guitar & acg & 637 & 535\\
    Electric guitar & elg & 760 & 942\\
    Organ & org & 682 & 361\\
    Piano & pia & 721 &995\\
    Saxophone & sax & 626 & 326\\
    Trumpet & tru & 577 &167\\
    Violin & vio & 580 & 211\\
    Voice & voi & 778 & 1044\\
    \hline
    \hline
\end {tabular}
\label{tab:instruments}
\end{table}

\subsection{Testing Configuration}

In the training phase, we used a fixed length window because the input data for ConvNet should be in a specific fixed shape. However, our testing audios had variable lengths between 5 s and 20 s, which were much longer than those of the training audio. Developing a system that can handle variable length of input data is valuable because music in real life varies in its length. We performed short-time analysis using overlapping windows to obtain local instrument information in the audio excerpts. Since an annotation exists per audio clip, we observed multiple sigmoid outputs and aggregated them to make a clip-wise decision. We tried two different strategies for the aggregation, which are the average and the normalized sum, which are referred as \textit{S1} and \textit{S2} throughout the paper, respectively. 

For \textit{S1}, we simply took an average of the sigmoid outputs class-wise (i.e., instrument-wise) over the whole audio clip and thresholded it without normalization. This method is intended to capture the existence of each instrument with its mean probability such that it might return the result without any detected instrument. For \textit{S2}, we first summed all sigmoid outputs class-wise over the whole audio excerpt and normalized the values by dividing them with the maximum value among classes such that the values were scaled to be placed between zero and one, followed by thresholding. This method is based on the assumption that humans perceive the ``predominant'' instrument in a more relatively scaled sense such that the strongest instrument is always detected and the existence of other instruments is judged by their relative strength compared to the most activate instrument.

Majority vote, one of the most common choices for a number of classification tasks, is not used in our system. Majority vote first predicts the classes for each analysis frame and the one with more vote wins. However, using this method for our task would result in disregarding accompaniment instruments, piano for example, because a music signal is composed of various musical instruments and usually the sounds are overlapped in time domain, and a presence of accompaniments are usually much weaker than that of voice or lead instruments. 

As our target is to identify an arbitrary number of predominant instruments in testing data, instruments with aggregated value over the threshold were all considered as predominant instruments. Using a higher value for the identification threshold will lead to better precision, but it will obviously decrease the recall. On the other hand, a lower threshold will increase the recall, but will lower the precision. Hence, we tried a range of values for the threshold to find the optimal value for the $F1$ measure, which is explained in the next Performance Evaluation section.

\begin{figure}[t!]
  \centering
  \includegraphics[keepaspectratio, width=3.5in]{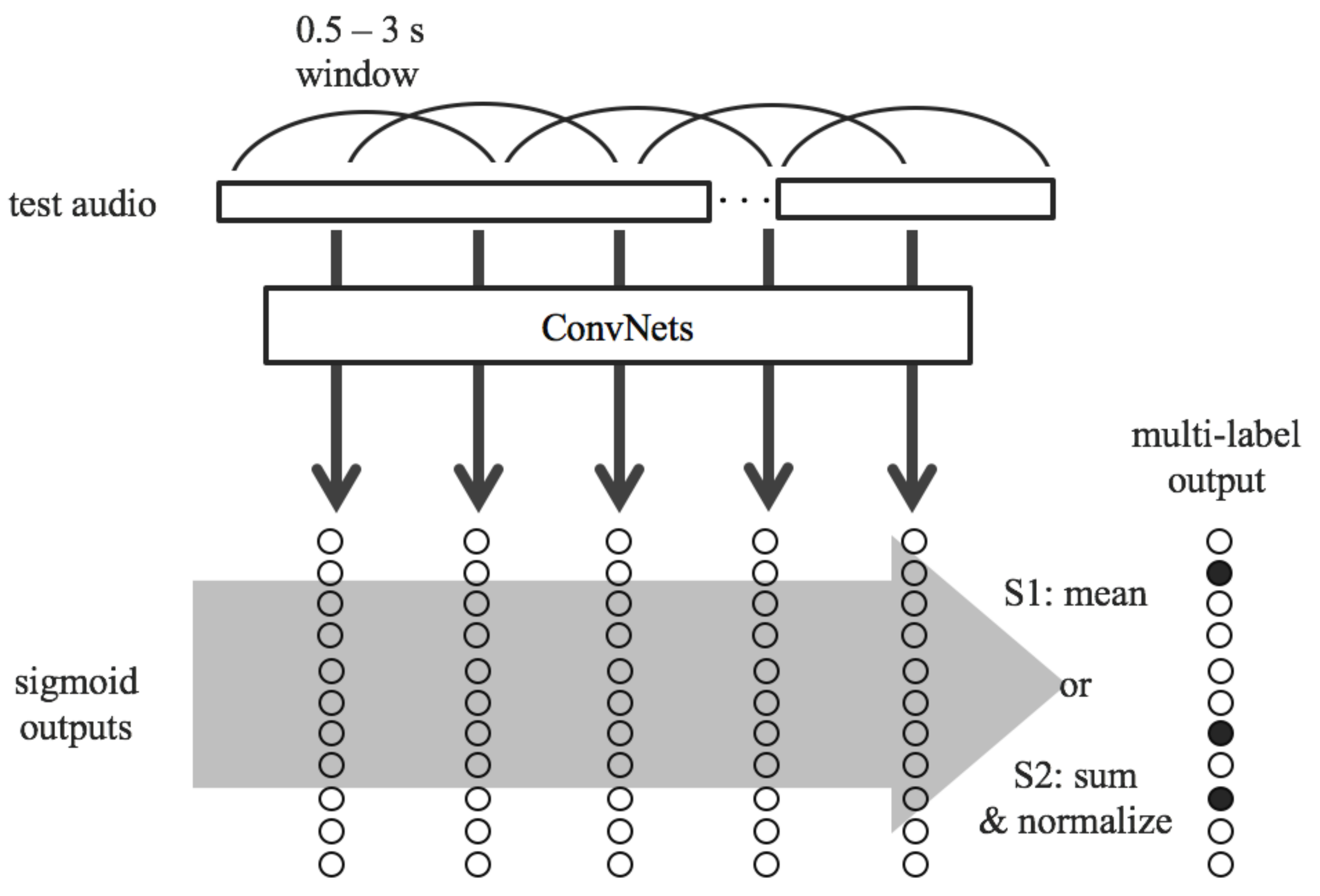}
  \caption{Schematic of obtaining a multi-label output from a test audio signal. Input audio was analyzed with sliding window, and these multiple sigmoid outputs were aggregated using two different strategies, \textit{S1} and \textit{S2}, to estimate the predominant instrument for the testing audio excerpt.}
\label{fig:schematic}
\end{figure}

For \textit{S1}, values between 0.02 and 0.18 were used, and for \textit{S2}, values between 0.2 and 0.6 were used as a threshold $\theta$. These threshold values were empirically chosen but set to be a wide enough range to find the best performance (i.e., highest $F1$ measure). The schematic of this aggregation process is illustrated in Figure \ref{fig:schematic}.

\subsection{Performance Evaluation}

Following the evaluation method widely used in the instrument recognition task, we computed the precision and recall, which are defined as

\begin{equation}
P = \frac{tp}{tp + fp}
\end{equation}

\begin{equation}
R = \frac{tp}{tp + fn}
\end{equation}

\noindent where $tp$ is true positive, $fp$ is false positive, and $fn$ is false negative. In addition, we used the $F1$ measure to calculate the overall performance of the system, which is the harmonic mean between precision and recall:

\begin{equation}
F1 = \frac{2PR}{P + R}
\end{equation}

Since the number of annotations for each class (i.e., 11 musical instruments) was not equal, we computed the precision, recall, and $F1$ measure for both the micro and the macro averages. For the micro averages, we calculated the metrics globally regardless of classes, thus giving more weight to the instrument with a higher number of appearances. On the other hand, we calculated the metrics for each label and found their unweighted average for the macro averages; hence, it is not related to the number of instances, but represents the overall performance over all classes. Finally, we repeated each experiment three times and calculated the mean and standard deviation of the output.

\begin{table}[!t]
\renewcommand{\arraystretch}{1.3}
\caption{Experiment Variables for the Activation Function, Size of the Analysis Window, Aggregation Strategy, and Identification Threshold. Default Settings are Indicated in Bold.}
\centering
\begin{tabular}{l | l}
    \hline
    \hline
    Variables\\
    \hline
    Activation func. & tanh, ReLU, PReLU, LReLU (0.01), \textbf{LReLU (0.33)}\\
    analysis win. size & 0.5 s, \textbf{1.0 s}, 1.5 s, 3.0 s\\
    Agg. strategy & \textit{S1} (mean), \textbf{\textit{S2} (sum and normalized)}\\
    $\theta$ (\textit{S1})& 0.02, 0.04, 0.06, 0.08, 0.10, 0.12, 0.14, 0.16, 0.18 \\
    $\theta$ (\textit{S2})& 0.20, 0.25, 0.30, 0.35, 0.40, 0.45, \textbf{0.50}, 0.55, 0.60 \\
    \hline
    \hline
\end {tabular}
\label{tab:expvar}
\end{table}

\section{Results}

We used LReLU ($\alpha$ = 0.33) for the activation function, 1 s for the analysis window, \textit{S2} for the aggregation strategy, and 0.50 for the identification threshold as default settings of the experiment where possible, which showed the best performance. The experiment variables are listed in Table \ref{tab:expvar}.

 First, we compared the performance of the proposed ConvNet with that of the existing algorithm on the IRMAS dataset. The effect of activation function, analysis window, aggregation strategy, and identification threshold on the recognition performance was analyzed separately in the following subsections.

\begin{figure}[!t]
  \centering
  \includegraphics[keepaspectratio, width=3.5in]{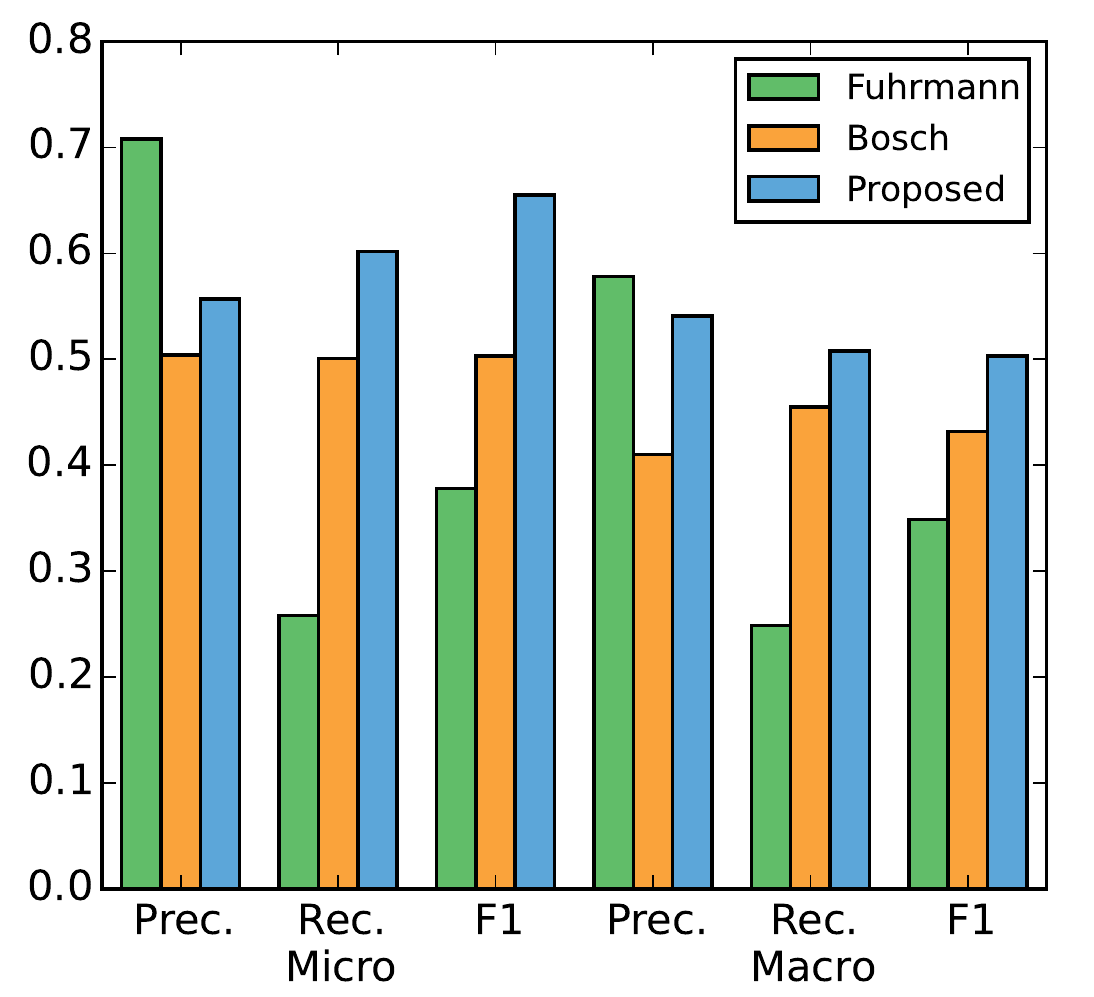}
  \caption{Performance comparison of the predominant instrument recognition algorithm from Fuhrmann and Herrra \cite{fuhrmann2010polyphonic}, Bosch et al. \cite{bosch2012comparison}, and our proposed ConvNet.}
\label{fig:perf_comp}
\end{figure}

\subsection{Comparison to Existing Algorithms}

For the result, our network achieved 0.602 for the micro $F1$ measure and 0.503 for the macro $F1$ measure. The existing algorithm from Fuhrmann and Herrera \cite{fuhrmann2010polyphonic} used typical hand-made timbral audio features with their frame-wise mean and variance statistics to train SVMs, and Bosch et al. \cite{bosch2012comparison} improved this algorithm with source separation called FASST (Flexible Audio Source Separation Framework) \cite{ono2010harmonic} in a preprocessing step. 

In terms of precision, Fuhrmann and Herrera's algorithm showed the best performance for both the micro and the macro measure. However, its recall was very low, around 0.25, which resulted in a low $F1$ measure. Our proposed ConvNet architecture outperformed existing algorithms on the IRMAS dataset for both the micro and the macro $F1$ measure, as shown in Figure \ref{fig:perf_comp}. From this result, it can be observed that the learned feature from the input data that is classified through ConvNet works better than the conventional hand-crafted features with SVMs.

\subsection{Effect of Activation Function}

In the case of using rectified units as an activation function, it was possible to observe a significant performance improvement compared to the tanh baseline as expected, as shown in Table \ref{tab:act_func_comp}. Unlike the result presented in the ImageNet classification work from He et al. \cite{he2015delving}, PReLU did not show any performance improvement, but just showed a matching performance with ReLU in our task. On the other hand, using LReLU showed better performance than using normal ReLU and PReLU. While using LReLU with a small gradient ($\alpha$ = 0.01) showed similar performance to ReLU as expected, LReLU with a very leaky alpha setting ($\alpha$ = 0.33) showed the best identification performance, which matched the result of the empirical evaluation work on ConvNet activation function from Xu et al. \cite{xu2015empirical}. 

This result shows that suppressing the negative part of the activation rather than making it all zero certainly improves the performance compared to normal ReLU because making the whole negative part zero might cause some initially inactive units to be never active as mentioned above. Moreover, this result shows that using leaky ReLU, which has been proved to work well in the image classification task, also benefits the musical instrument identification.

\begin{figure}[t!]
  \centering
  \includegraphics[keepaspectratio, width=3.5in]{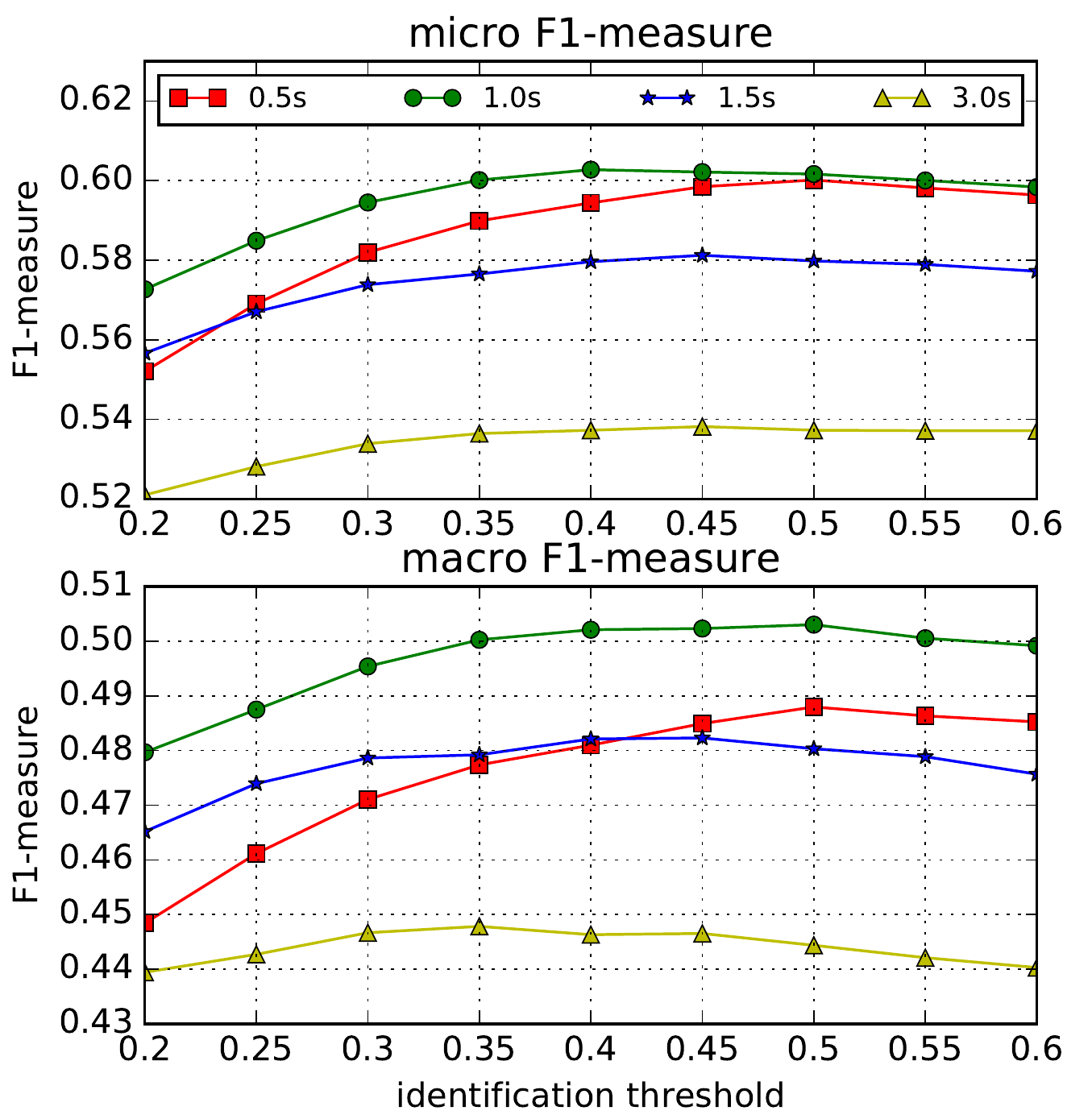}
  \caption{Micro and macro $F1$ measure of an analysis window size of 0.5, 1.0, 1.5, and 3.0 s according to the identification threshold.}
\label{fig:win_size}
\end{figure}

\begin{table}[!t]
\renewcommand{\arraystretch}{1.3}
\caption{Instrument Recognition Performance of the Proposed ConvNet with Various Activation Functions.}
\centering
\begin{tabular}{l | l l l l l l}
    \hline
    \hline
    \multirow{2}{*}{Activation func.} & \multicolumn{3}{c}{Micro} & \multicolumn{3}{c}{Macro} \\
    & $P$ & $R$ & $F1$ & $P$ & $R$ & $F1$ \\
    \hline
    tanh & 0.416 & 0.625 & 0.499 & 0.348 & 0.537 & 0.399 \\
    ReLU & 0.640 & 0.550 & 0.591 & 0.521 & 0.508 & 0.486 \\
    PReLU & 0.612 & 0.565 & 0.588 & 0.502 & 0.516 & 0.490 \\
    LReLU ($\alpha$=0.01) & 0.640 & 0.552 & 0.593 & 0.530 & 0.507 & 0.492 \\
    LReLU ($\alpha$=0.33)  & 0.655 & 0.557 & 0.602 & 0.541 & 0.508 & 0.503 \\
    \hline
    \hline
\end {tabular}
\label{tab:act_func_comp}
\end{table}

\subsection{Effect of Analysis Window Size}

As mentioned above, we conducted an experiment with diverse analysis window sizes such as 3.0, 1.5, 1.0, and 0.5 s to find the optimal analysis resolution. Figure \ref{fig:win_size} shows the micro and macro $F1$ measure with various analysis frame sizes according to identification threshold, and it can be observed that the use of the longest 3.0-s window clearly performed poorer than the use of shorter window sizes regardless of identification threshold. However, shortening the analysis frame down to 0.5 s decreased the overall performance again.

From this result, it can be seen that 1.0 s is the optimal analysis window size for our task. Using a shorter analysis frame helped to increase the temporal resolution, but 0.5 s was found to be too short a window size for identifying the instrument.

\begin{figure*}[!t]
  \centering
  \includegraphics[keepaspectratio, width=\textwidth]{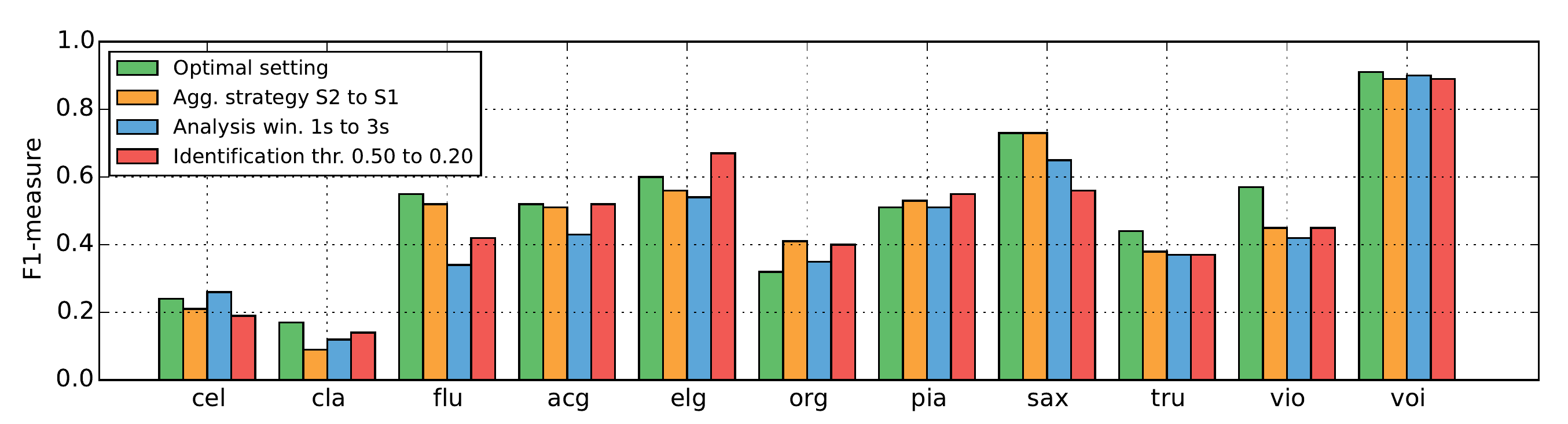}
  \caption{Class-wise performance of the instrument identification. To analyze the effect of each parameter on each instrument, we compared the optimal setting (i.e., default setting) with a different aggregation strategy, analysis window size, and LReLU identification threshold.}
\label{fig:class_wise}
\end{figure*}

\subsection{Effect of Identification Threshold}

Using a higher value for the identification threshold leads to better precision, but it decreases the recall. On the contrary, a lower threshold results in better recall with lower precision. Hence, we used the $F1$ measure, which is the harmonic mean of precision and recall to evaluate the overall performance. In terms of $F1$ measure, we found that 0.5 is the most appropriate threshold as it showed the best performance for the macro $F1$ measure, as shown in Figure \ref{fig:win_size}. 

The current system uses a certain identification threshold for all instruments. However, we think that there might be a room for improvement by using different thresholds for each instrument because there are various types of instruments included in the experiment. For example, the amplitude of the piano sound was relatively small in a number of music excerpts because it is usually used as an accompanying instrument. On the other hand, the flute sound in the music was mostly louder than others because it is usually used as a lead instrument.

\begin{table}[!t]
\renewcommand{\arraystretch}{1.3}
\caption{Instrument Recognition Performance of the Proposed ConvNet Using Two Different Aggregation Strategies with the Threshold $\theta$ that Returned the Highest $F1$ Measure for Each Strategy for Comparison.}
\centering
\begin{tabular}{l | l l l l l l}
    \hline
    \hline
    \multirow{2}{*}{Agg. strategy} & \multicolumn{3}{c}{Micro} & \multicolumn{3}{c}{Macro} \\
    & $P$ & $R$ & $F1$ & $P$ & $R$ & $F1$ \\
    \hline
    S1 ($\theta$ = 0.16)& 0.627 & 0.572 & 0.595 & 0.525 & 0.502 & 0.486 \\
    S2 ($\theta$ = 0.50)& 0.655 & 0.557 & 0.602 & 0.541 & 0.508 & 0.503 \\
    \hline
    \hline
\end {tabular}
\label{tab:theta_comp}
\end{table}

\subsection{Effect of Aggregation Strategy}
We conducted an experiment with two different strategies, \textit{S1} and \textit{S2}, for the aggregation of ConvNet outputs as explained in the Testing Configuration section. The performance of \textit{S1} and \textit{S2} is demonstrated in Table \ref{tab:theta_comp} with a threshold $\theta$ that returned the highest $F1$ measure for each strategy. As a result, \textit{S2} showed better identification performance than \textit{S1} overall. There was only a slight performance gap between \textit{S1} and \textit{S2} for the micro $F1$ measure, but the difference was notable for the macro $F1$ measure. This result shows that performing a class-wise sum followed by normalization is a better aggregation method for predominant instrument identification than taking class-wise mean values. It is likely due to the training and testing audios differing in quality to a great extent, depending on the recording and production time, and the audio-excerpt-wise normalization helped to minimize the effect of quality differences between audio excerpts, which would result in a more generalized output.

\subsection{Analysis of Instrument-Wise Identification Performance}

The results demonstrated above were focused on the overall identification performance. In this section, we analyze and discuss the result instrument-wise (i.e., class-wise) to observe the system performance in detail. As shown in Figure \ref{fig:class_wise}, identification performance varies to a great extent, depending on the instruments. Regardless of parameter setting, it can be observed that the system recognizes the voice in the music very well, showing an $F1$ measure of about 0.90. On the other hand, cello and clarinet showed relatively poor performance compared to other instruments, showing an $F1$ measure of around 0.20. 

These results were highly likely affected by the insufficient number of training audio samples. For deep learning, the number of training examples is critical for the performance compared to the case of using hand-crafted features because it aims to learn a feature from the low-level input data. As illustrated in Table \ref{tab:instruments}, the number of training audio samples for voice is 778, which is the largest number of training audio. On the contrary, 338 and 505 audio excerpts were used for cello and clarinet, respectively, which were the least and third least number of training audio. We believe that increasing the number of training data for cello and clarinet would be helpful to increase the identification performance for these instruments.

In addition, the number of test audio samples for cello and clarinet was much less than those for other instruments too. The dataset only has 111 and 62 test audio samples for cello and clarinet, respectively, which are the first and second least number of test audio, while it has 1044 audio samples for the human voice. Evaluating the system on a small number of test data would make the result less reliable and less stable than other identification results.

\begin{figure*}[!t]
  \centering
  \includegraphics[keepaspectratio, width=\textwidth]{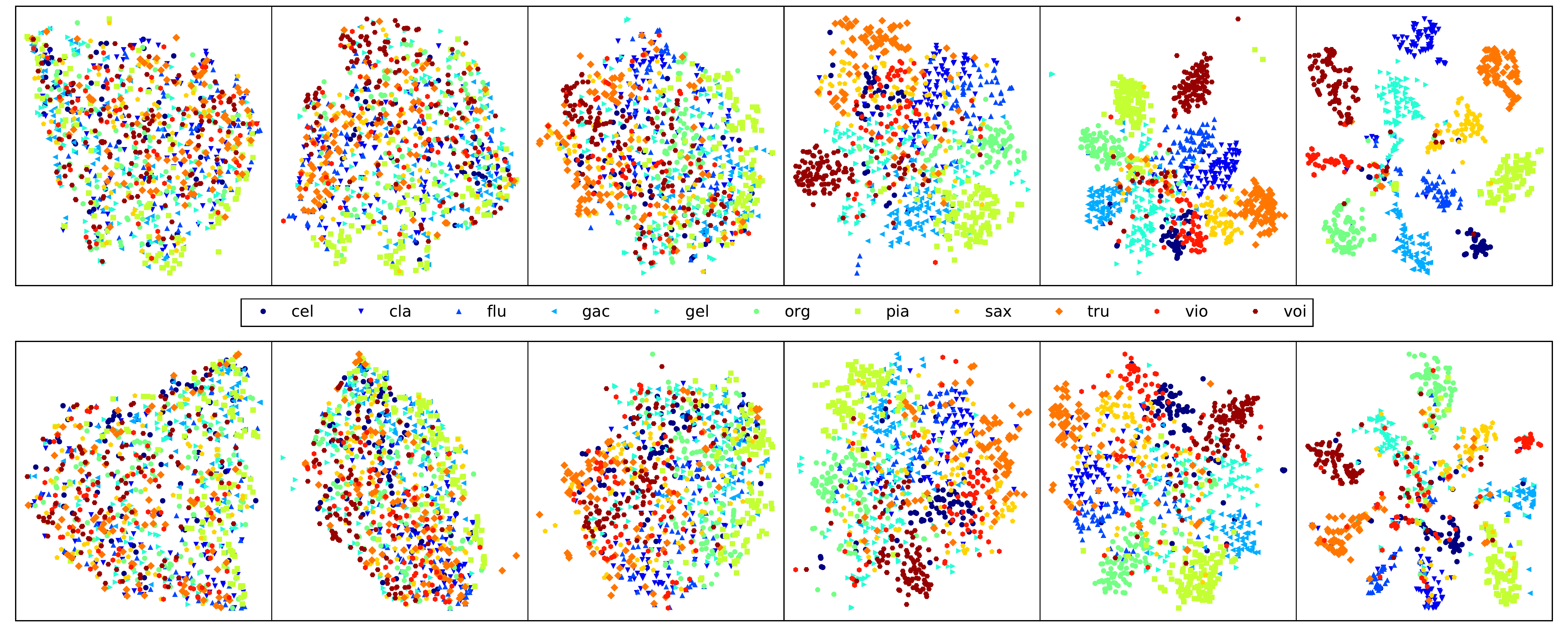}
  \caption{Visualization of the t-SNE clustering result. It represents the clustering result for each intermediate state of the proposed model. From left to right, the first four plots are the clustering result of the activations at the end of each convolutional block, and the last two plots are the clustering result of the activations of the hidden dense layer and the final sigmoid output, respectably. The upper plots are drawn from the sample used in the training, and the lower plots are from the validation data samples.}
\label{fig:cluster_tsne_result_1s}
\end{figure*}

Apart from the issue related to the number of audio, high identification performance of the voice class is highly likely owing to its spectral characteristic that is distinct from other musical instruments. The other instruments used in the experiment usually produce relatively clear harmonic patterns; however, the human voice produces highly unique spectral characteristics that contain much more inharmonic spectrum with a natural vibrato.

Regarding aggregation strategy, using \textit{S1} instead of \textit{S2} decreased the identification performance for organ and piano. This result indicates that \textit{S1} showed a slight advantage on instruments that are usually used as an accompaniment instrument, while using \textit{S1} for aggregation was better for most of the cases. On the other hand, using a 3-s analysis window instead of the default 1-s window considerably decreased the performance, especially for flute, acoustic guitar, electric guitar, and violin. This result shows that using a longer analysis window is a disadvantage for most of the cases. Finally, using a very low identification threshold, 0.20, caused considerable performance loss especially for flute, saxophone, trumpet, and violin, while it showed a slight improvement for electric guitar, organ, and piano. 

This result can be understood to mean that using a lower threshold for identification performance helps to detect instruments that are usually in the background, while using a higher threshold is suitable for instruments that are frequently used as a lead instrument or for wind instruments that usually show relatively strong presence in the music. As mentioned in the Results section, this result indicates that there can be a potential performance improvement by using a different identification threshold for each instrument.

\subsection{Qualitative Analysis with Visualization Methods}

To understand the internal mechanism of the proposed model, we conducted a visual analysis with various visualization methods. First, we tried clustering for each layer's intermediate hidden states from a given input data sample to verify how the encoding behavior of each layer contributes to the clustering of input samples. We selected the t-distributed stochastic neighbor embedding (t-SNE) \cite{van2008visualizing} algorithm, which is a technique for dimensionality reduction of high-dimensional data. Second, we exploited the deconvolution \cite{zeiler2014visualizing,yosinski2015understanding} method to identify the functionality of each unit in the proposed ConvNet model by visual analysis. Our system basically repeats two convolutional layers followed by one pooling layer, and we grouped these three components and call it ``convolutional block'' throughout this section for simplicity.

The t-SNE algorithm is based on the stochastic neighbor embedding (SNE) algorithm, which converts the similarities between given data points to joint probability and then embeds high-dimensional data points to lower-dimensional space by minimizing the Kuller-Leibler divergence between the joint probability of low-dimensional embedding and the high-dimensional data points. This method is highly effective especially in a dataset where its dimension is very high \cite{van2008visualizing}. This advantage of the algorithm accorded well with our condition, where the target observations were necessarily in a high dimension since we reshaped each layer's filter activations to a single vector respectively.

With the visualization exploiting t-SNE, we could observe how each layer contributed to the classification of the dataset. Reflecting a gradually changing inter-distance of data points at each stage of the proposed model, four intermediate activations were extracted at the end of each convolutional block and one from the hidden fully connected layer, and another one from the final output layer. For the compression of dimensionality and computational efficiency, we pooled the maximum values for activation matrices of each unit. By this process, the dimensionality of each layer's output could be diminished to each layer's unit size. We visualized on both randomly selected training and validation data samples from the entire dataset to verify both how the model exactly works and how it generalizes its classification capability. In Figure \ref{fig:cluster_tsne_result_1s}, it is clearly shown that data samples under the same class of instrument are well grouped and each group is separated farther, with the level of encoding being higher, particularly on the training set. While the clustering was not clearer than the former case, the tendency of clustering on the validation set was also found to be similar to the training set condition.


\begin{figure*}[!t]
  \centering
  \includegraphics[keepaspectratio, width=\textwidth]{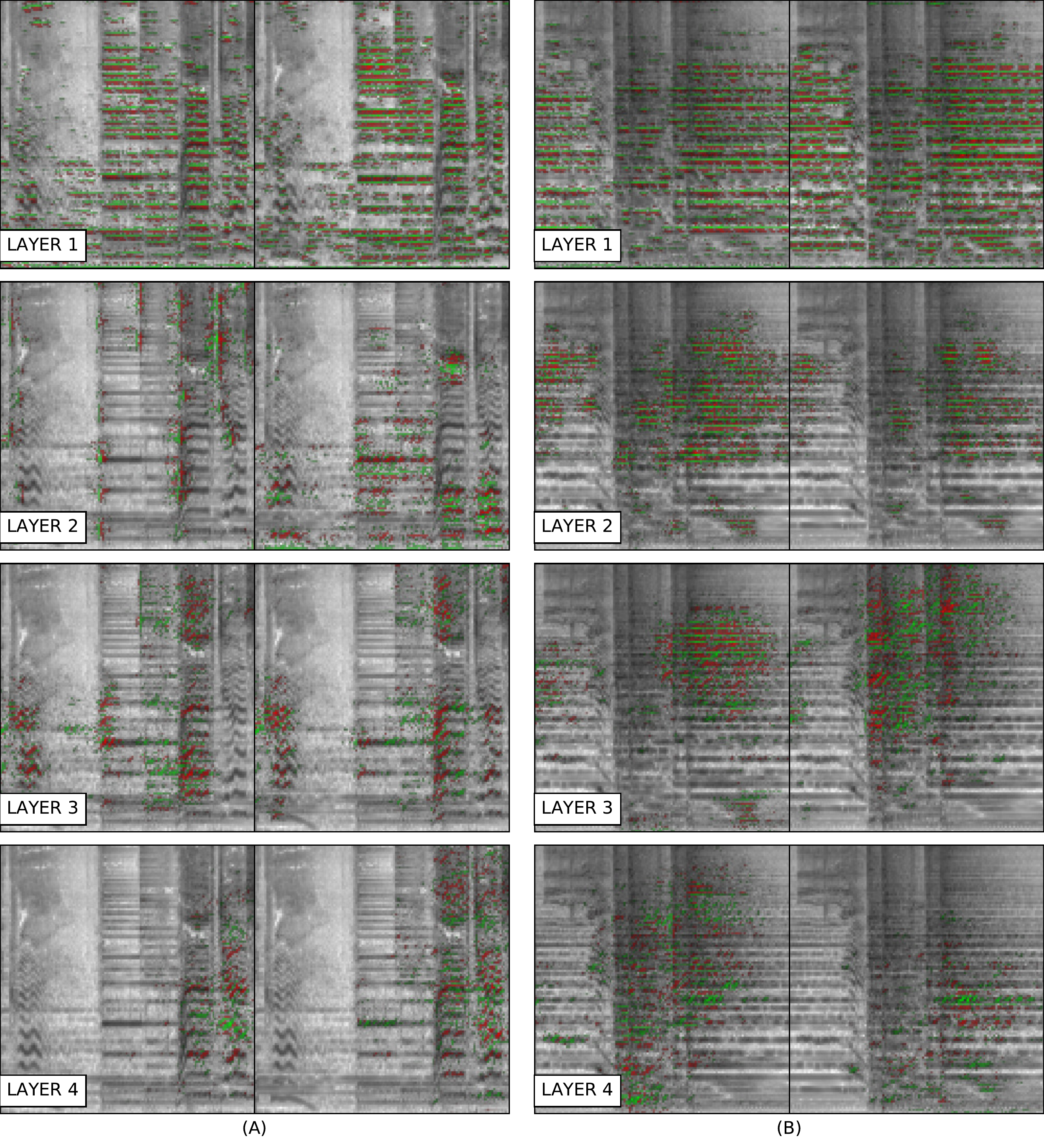}
  \caption{Mel-spectrogram of two input signals and their respective deconvoluted results. The left two columns and right two columns of the image, denoted as (A) and (B), respectively, were calculated from two independent music signals. Both signals were a 3-s polyphonic music segment that was randomly cropped from the original music. Moreover, both signals (A) and (B) consist mainly of the voice and the acoustic guitar sound. However, the dominant instrument of (A) is labeled as the voice, while (B) is labeled as the acoustic guitar. Each row of images represents a deconvoluted signal overlaid on the original signal. We extracted these results from four intermediate stages of the proposed model. Deconvolution outputs were extracted from the end of each convolutional block. For both target signals, the two highest activated units of each point were chosen and deconvoluted to be visualized. From left to right, images are arranged in order of decreasing absolute unit activation. The red region and the green region of each deconvoluted image indicate the positive value and negative value of the result, respectably. The remaining area is where the magnitude of activation is relatively lower than in those regions. The range of activation result is normalized for the purpose of clear visualization.}
\label{fig:DECONV_RES_NEW}
\end{figure*}


Another visualization method, deconvolution, has recently been introduced as a useful analysis tool to qualitatively evaluate each node of a ConvNet. The main principle of this method is to inverse every stage of operations reaching to the target unit, to generate a visually inspectable image that has been, as a consequence, filtered by the trained sub-functionality of the target unit \cite{zeiler2014visualizing}. With this method, it is possible to reveal intuitively how each internal sub-function works within the entire deep convolutional network, which tends to be thought of as a ``black box''.

By this process, the functionality of a sub-part of the proposed model is explored. We generated deconvoluted images like those in Figure \ref{fig:DECONV_RES_NEW} from the arbitrary input mel-spectrogram, for each unit in the entire model. From the visual analysis of the resulting images, we could see several aspects of the sub-functionalities of the proposed model: (1) Most units in the first layer tend to extract vertical, horizontal, and diagonal edges from the input spectrogram, just like the lower layers of ConvNets do in the usual image object recognition task. (2) From the second layer through the fourth layer, each deconvoluted image indicates that each unit of the mid-layers has a functionality that searches for particular combinations of the edges extracted from the first layer. (3) It was found that it is difficult to strongly declare each sub-part of the proposed model that detects a specific musical articulation or expression. However, in an inductive manner, we could see that some units indicate that they can be understood as a sub-function of such musical expression detector.

We conducted a visual analysis of the deconvoluted image of two independent music signals, which have the same kind of sound sources, but differently labeled.\footnote{Both signals were composed of a ``voice'' and an ``acoustic guitar'' instrument, but the predominant instrument of signal (A) was labeled as the ``voice,'' while (B) was labeled as the ``acoustic guitar''.} For both cases, the most activated units of the first layer strongly suggested that their primary functionality is to detect a harmonic component on the input mel-spectrogram by finding horizontal edges in it, as shown in the top figures in Figure \ref{fig:DECONV_RES_NEW}. However, from the second layer to higher layers, the highly activated units' behavior appeared to be quite different for each respective input signal. For instance, the most activated unit of signal (A)'s second layer showed a functionality similar to onset detection, by detecting a combination of vertical and horizontal edges. Compared to this unit, the most activated units of the third layer showed a different functionality that seems to activate unstable components such as the vibrato articulation or the ``slur'' of the singing voice part, by detecting a particular combination of diagonal and horizontal edges. On the other hand, the model's behavior in signal (B) was very different. As is clearly shown in the second and the third layers' output in Figure \ref{fig:DECONV_RES_NEW}, the highly activated sub-functions were trying to detect a dense field of stable, horizontal edges which are often found in harmonic instruments like guitar. Each field detected from those units corresponded to the region where the strumming acoustic guitar sound is.



\section{Conclusion}

In this paper, we described how to apply ConvNet to identify predominant instrument in the real-world music. We trained the network using fixed-length single-labeled data, and identify an arbitrary number of the predominant instrument in a music clip with a variable length.

Our results showed that very deep ConvNet is capable of achieving good performance by learning the appropriate feature automatically from the input data. Our proposed ConvNet architecture outperformed previous state-of-the-art approaches in a predominant instrument identification task on the IRMAS dataset. Mel-spectrogram was used as an input to the ConvNet, and we did not use any source separation in the preprocessing unlike in existing works.

We conducted several experiments with various activation functions for ConvNet. Tanh and ReLU were used as a baseline, and the recently introduced LReLU and PReLU were also evaluated. Results confirmed that ReLU worked reasonably well, which is a de facto standard in recent ConvNet studies. Furthermore, we obtained the better results with LReLU than with normal ReLU, especially with the very leaky setting ($\alpha$ = 0.33). The performance of Tanh was worse than those of other rectifier functions as expected, and PReLU just showed a matching performance with ReLU for our task. 

This paper also investigated different aggregation methods for ConvNet outputs that can be applied to music excerpts with various lengths. We experimented with two different aggregation methods, which are the class-wise mean probability \textit{S2} and the class-wise sum followed by normalization \textit{S2}. The experimental results showed that \textit{S2} is a better aggregation method because it effectively deals with the quality difference between audios through the audio-excerpt-wise normalization process. In addition, we conducted an extensive experiment with various analysis window sizes and identification thresholds. For the analysis window size, using a shorter window improved the performance by increasing the temporal resolution. However, 0.5 s was too short to obtain an accurate identification performance, and 1.0 s was found to be the optimal window size. There was a trade-off between precision and recall, depending on the identification threshold; hence, we used an $F1$ measure, which is the harmonic mean of precision and recall. For the result, a threshold value of 0.5 showed the best performance.

Visualization of the intermediate outputs using t-SNE showed that the feature representation became clearer each time the input data were passed through the convolutional blocks. Moreover, visualization using deconvolution showed that the lower layer tended to capture the horizontal and vertical edges, and that the higher layer tended to seek the combination of these edges to describe the spectral characteristics of the instruments.

Our study shows that many recent advances in a neural network on the image processing area are transferable to the audio processing domain. However, audio signal processing, especially music signal processing, has many different aspects compared to the image processing area where ConvNets are most extensively used. For example, spectral characteristics are usually overlapped in both time and frequency unlike the objects in an image, which makes the detection difficult. Moreover, music signals are much more repetitive and continuous compared to natural images and are present in various lengths. We believe that applying more musical knowledge on the aggregation part with adaptive thresholding for each instrument can improve the performance further, which warrants deeper investigation.

\section*{Acknowledgment}
This research was supported partly by the MSIP (Ministry of Science, ICT and Future Planning), Korea, under the ITRC (Information Technology Research Center) support program (IITP-2016-H8501-16-1016) supervised by the IITP (Institute for Information \& communications Technology Promotion), and partly by a National Research Foundation of Korea (NRF) grant funded by the MSIP (NRF-2014R1A2A2A04002619).

\ifCLASSOPTIONcaptionsoff
  \newpage
\fi



%

\bibliographystyle{IEEEtran}


%

\begin{IEEEbiography}[{\includegraphics[width=1in,height=1.25in,clip,keepaspectratio]{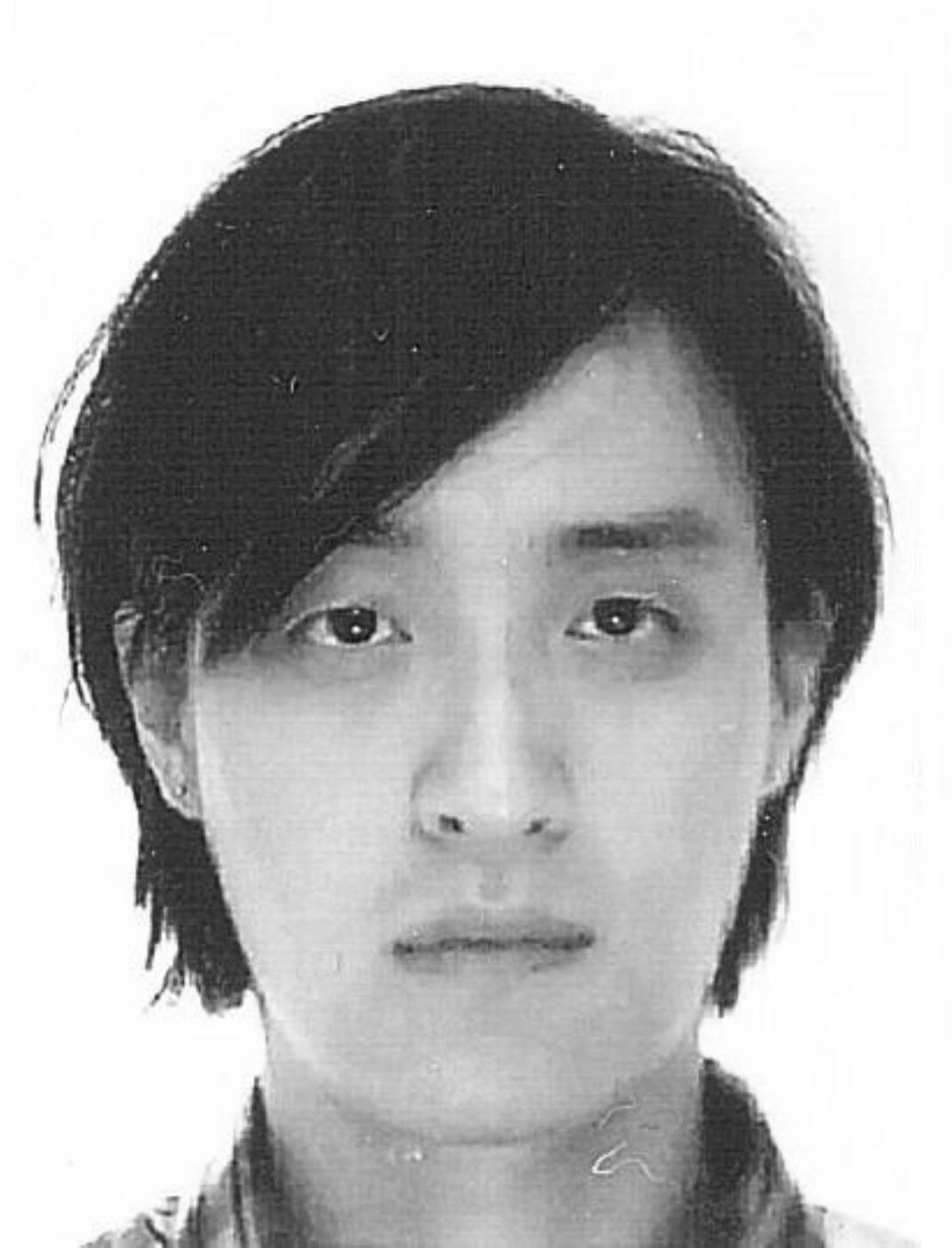}}]{Yoonchang Han}
was born in Seoul, Republic of Korea, in 1986. He studied electronic engineering systems at King's College London, UK, from 2006 to 2009, and then moved to Queen Mary University of London, UK, and received an MEng (Hons) degree in digital audio and music system engineering with First Class Honours in 2011. He is currently a PhD candidate in digital contents and information studies at the Music and Audio Research Group (MARG), Seoul National University, Republic of Korea. His main research interest lies within developing deep learning techniques for automatic musical instrument recognition.
\end{IEEEbiography}

\begin{IEEEbiography}[{\includegraphics[width=1in,height=1.25in,clip,keepaspectratio]{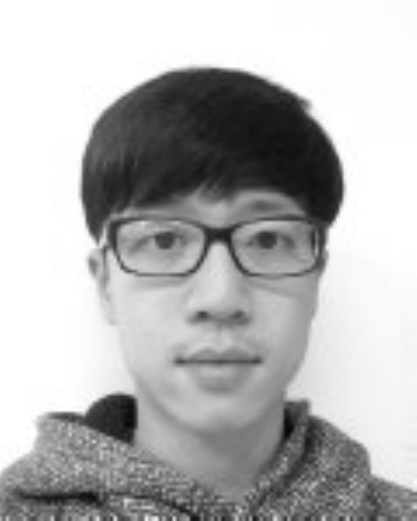}}]{Jaehun Kim}
was born in Seoul, Republic of Korea, in 1986. He is currently a researcher at the Music and Audio Research Group. His research interests include signal processing and machine learning techniques applied to music and audio. He received a BA in English literature and linguistics from Seoul National University and received an MS degree in the digital contents and information studies at the Music and Audio Research Group (MARG), Seoul National University, Republic of Korea.
\end{IEEEbiography}

\begin{IEEEbiography}[{\includegraphics[width=1in,height=1.25in,clip,keepaspectratio]{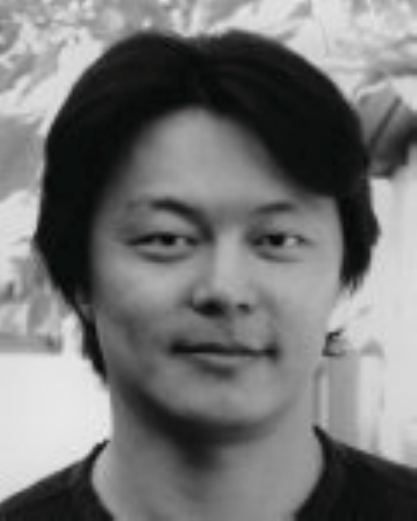}}]{Kyogu Lee}
is an associate professor at Seoul National University and leads the Music and Audio Research Group. His research focuses on signal processing and machine learning techniques applied to music and audio. Lee received a PhD in computer-based music theory and acoustics from Stanford University.
\end{IEEEbiography}


\vfill


\end{document}